\documentclass[epj]{webofc}
\usepackage[varg]{txfonts}   % Web of Conferences font
\newcommand{\be}{\begin{equation}}
\newcommand{\ee}{\end{equation}}
\newcommand{\bea}{\begin{eqnarray}}
\newcommand{\eea}{\end{eqnarray}}
\wocname{New Frontiers in Physics}
\woctitle{New Frontiers in Physics 2014}
\begin{document}
\title{Electromagnetic probes of the QGP}

\author{E. L. Bratkovskaya\inst{1}\fnsep\thanks{\email{Elena.Bratkovskaya@th.physik.uni-Frankfurt.de}} \and
        O. Linnyk\inst{2} \and
        W. Cassing\inst{2}    }

\institute{Institut f\"ur Theoretische Physik,
 Johann Wolfgang Goethe University Frankfurt/M,
 Germany
\and Institut f\"ur Theoretische Physik, University of Giessen,
Germany
          }

\abstract{We investigate the properties of the QCD matter across the
  deconfinement phase transition in the scope of the parton-hadron string
  dynamics (PHSD) transport approach. We present here in particular the
  results on the electromagnetic radiation, i.e. photon and dilepton
  production, in relativistic heavy-ion collisions. By comparing our calculations
for the heavy-ion collisions to the available data, we determine the relative
importance of the various production sources and address the
possible origin of the observed strong elliptic flow $v_2$ of direct
photons. We argue that the different centrality dependence of the hadronic
and partonic sources for direct photon production in nucleus-nucleus
collisions can be employed to shed some more light on the origin of the
 photon $v_2$ "puzzle".   While the dilepton spectra at low invariant mass show
in-medium effects like an enhancement from multiple baryonic
resonance formation or a collisional broadening of the vector meson
spectral functions, the dilepton yield at high invariant masses
(above 1.1 GeV)  is dominated by QGP contributions for central
heavy-ion collisions at ultra-relativistic energies. This allows to have
an independent view on the parton dynamics via their electromagnetic massive radiation.}
\maketitle

\section{Introduction}
\label{intro}
The electromagnetic emissivity of strongly interacting matter is a
subject of longstanding interest \cite{FeinbShur,ChSym}
and is explored in particular in
relativistic nucleus-nucleus collisions, where the photons (and
dileptons) measured experimentally provide a time-integrated picture
of the collision dynamics.  The recent observation by the PHENIX
Collaboration~\cite{PHENIX2010} that the elliptic flow $v_2(p_T)$ of
'direct photons' produced in minimal bias Au+Au collisions at
$\sqrt{s_{NN}}=200$~GeV is comparable to that of the produced pions
was a surprise and in contrast to the theoretical expectations and
predictions. We will analyse this photon $v_2$ "puzzle" within
scope of the parton-hadron string dynamics (PHSD) transport approach \cite{PHSD}
with a focus on the centrality dependence of the different production sources.
Furthermore, the PHSD approach will be used to study dilepton production
in nucleus-nucleus collisions from SIS to LHC energies in comparison to available data
in order to extract information about
the modification of hadron properties in the dense and hot hadronic
medium which might shed some light on chiral symmetry restoration
(cf. \cite{ChSym} and references therein). On the other hand we intend to
identify those spectral regimes where we see a clear dominance of partonic channels
that might allow to determine their transport properties via their
electromagnetic emissivity.

%----------------------------------------------------------------------------------
\section{Photon/dilepton emission rates}

\noindent In hydrodynamical calculations for the time evolution of
the bulk matter the equilibrium emission rate of electromagnetic
probes enters which in thermal field theory can be expressed as \cite{Rate1,Rate2}: \\
\begin{eqnarray}
q_0 {d^3R\over d^3q}= -{\dfrac{g_{\mu\nu}}{(2\pi)^3}}
       Im \Pi^{\mu\nu} (q_0=|\vec{q}|) f(q_0,T);
\label{RatePh}
\end{eqnarray}
for photons with 4-momentum $q=(q_0,\vec q)$ and
\begin{eqnarray}
E_+E_- {d^3R \over {d^3p_+d^3p_-}}= \dfrac{2e^2}{(2\pi)^6} \dfrac{1}{q^4}
       L_{\mu\nu} Im \Pi^{\mu\nu} (q_0,|\vec{q}|)f(q_0,T).
\label{RateDil}
\end{eqnarray} for dilepton pairs
with 4-momentum  $q=(q_0,\vec q)$, where $q=p_+ + p_-$ and
$p_+=(E_+,\vec p_+), p_-=(E_-,\vec p_-)$. Here the Bose distribution
function is $f(q_0,T) =1/(e^{q_0/T}-1)$; $L_{\mu\nu}$ is the
electromagnetic leptonic tensor, $\Pi^{\mu\nu}$ is the retarded
photon self-energy at finite temperature $T$ related to the
electromagnetic current correlator $\Pi^{\mu\nu} \sim i\int d^4x
e^{ipx}\left\langle[J_\mu(x),J_\nu(0)]\right\rangle|_T $. Using the
Vector-Dominance-Model (VDM) $Im\Pi^{\mu\nu}$ can be related to the
in-medium $\rho$-meson spectral function from many-body approaches
\cite{RappWam97} which, thus, can be probed by dilepton measurements
directly. The photon rates for $q_0\to 0$ are related to the
electric conductivity $\sigma_0$ which allows to probe the electric
properties of the QGP \cite{Cass13PRL}. We point out  that
Eqs.(\ref{RatePh}),(\ref{RateDil}) are strictly applicable only for
systems in thermal equilibrium whereas the dynamics of heavy-ion
collisions is generally of non-equilibrium nature.

\noindent The non-equilibrium emission rate from relativistic
kinetic theory \cite{Rate2,RateKT2}, e.g. for the process $1+2\to
\gamma +3$, is
\begin{equation}
q_0 {d^3R\over d^3q}= \int \dfrac{d^3p_1}{2(2\pi)^3 E_1}
\dfrac{d^3p_2}{2(2\pi)^3 E_2} \dfrac{d^3p_3}{2(2\pi)^3 E_3} \
(2\pi)^4 \ \delta^4(p_1+p_2-p_3-q) \ |M_{if}|^2 \
\dfrac{f(E_1) f(E_2) (1\pm f(E_3))}{2(2\pi)^3},
\label{RateKT}
\end{equation}
where $f(E_i)$ is the distribution function of particle $i=1,2,3$,
which can be hadrons (mesons and baryons) or partons. In Eq.
(\ref{RateKT}) $M_{if}$ is the matrix element of the reaction which
has to be evaluated on a microscopical level. In the case of
hadronic reactions One-Boson-Exchange models or chiral models are
used to evaluate $M_{if}$  on the level of Born-type diagrams.
However, for a consistent consideration of such elementary process
in the dense and hot hadronic environment, it is important to
account for the in-medium modification of hadronic properties, i.e.
many-body approaches such as self-consistent $G$-matrix calculations
have to be applied (e.g. \cite{Gmatrix} for anti-kaons or
\cite{RappWam97} for $\rho$ mesons).

%----------------------------------------------------------------------------------
\section{Photons}
\subsection{Production sources}

There are different production sources of photons in $p+p$ and $A+A$ collisions:\\
1) {\it Decay photons } -  most of the photons seen in $p+p$ and
$A+A$ collisions stem from the hadronic decays:
$m \to \gamma + X,  m = \pi^0, \eta, \omega, \eta^\prime, a_1, ...$. \\
2) {\it Direct photons} - obtained by subtraction of the decay photon contributions from
the inclusive (total) spectra measured experimentally.\\
(i) The are a few sources of direct photons at large transverse
momentum $p_T$ denoted by {\it 'hard'} photons: the 'prompt'
production from the initial hard $N+N$ collisions and the photons
from the jet fragmentation reactions, which are the standard pQCD
type of processes. The latter, however, might be modified in $A+A$
contrary to $p+p$ due to the parton energy loss in the medium. \\
(ii)
At low $p_T$ the photons come from the thermalized QGP, so called {\it 'thermal'} photons,
as well as from {\it hadronic} interactions: \\
$\bullet$
The {\it 'thermal'} photons from the QGP arise mainly from $q\bar q$ annihilation
($q+\bar q \to g +\gamma$) and Compton scattering ($q(\bar q) + g \to q(\bar q) + \gamma$)
which can be calculated in leading order pQCD  \cite{AMY01}.
However, the next-to-leading order corrections turn out to be also important \cite{JacopoQM14}.\\
$\bullet$
{\it Hadronic} sources of photons are related to \\
1) secondary mesonic interactions as $\pi + \pi \to \rho + \gamma, \
\rho + \pi \to \pi + \gamma, \ \pi + K \to \rho + \gamma, ....$ The
binary channels with $\pi, \rho$ have been evaluated in effective
field theory \cite{Kapusta91} and are used in transport model
calculations \cite{HSD08,Linnyk:Photon} within the extension for the
off-shellness of $\rho$-mesons due to the broad spectral function.
Alternatively, the binary hadron rates (\ref{RateKT}) have been
derived in the massive Yang-Milles approach in Ref. \cite{TRG04} and
been often used in hydro calculations . \\
2) hadronic bremsstrahlung, such as meson-meson ($mm$) and
meson-baryon ($mB$) bremsstrahlung $m_1+m_2\to m_1+m_2+\gamma, \ \
m+B\to m+B+\gamma$, where $m=\pi,\eta,\rho,\omega,K,K^*,...$ and
$B=p,\Delta, ...$. Here the leading contribution corresponds to the
radiation from one charged hadron. The importance of bremsstrahlung
contributions to the photon production will be discussed below.

%----------------------------------------------------------------------------------
\subsection{Direct photons and the $v_2$ 'puzzle'}

The photon production has been measured early in relativistic
heavy-ion collisions  by the WA98 Collaboration in S+Au and Pb+Pb
collisions at SPS energies  \cite{WA98}. The model comparisons with
experimental data show that the high $p_T$ spectra are dominated by
the hard 'prompt' photon production whereas the 'soft' low $p_T$
spectra stem from hadronic sources since the thermal QGP radiation
at SPS energies is not large. Moreover, the role of hadronic
bremsstrahlung turns out to be very important for a consistent
description of the low $p_T$ data as has been found a couple of
years ago in expanding fireball model calculations \cite{LiuRapp07}
and in the HSD (Hadron-String-Dynamics) transport approach
\cite{HSD08}. Unfortunately, the accuracy of the experimental data
at low $p_T$ did not allow to draw further solid conclusions.

The measurement of photon spectra by the PHENIX Collaboration
\cite{PHENIX2010} has stimulated a new wave of interest for direct
photons from the theoretical side since at RHIC energies the thermal
QGP photons have been expected to dominate the spectra. A variety of
model calculations based on fireball, Bjorken hydrodynamics, ideal
hydrodynamics with different initial conditions and
Equations-of-State (EoS) turned out to show substantial differences
in the slope and magnitude of the photon spectra (for a model
comparison see Fig. 47 of \cite{PHENIX2010} and corresponding
references therein). Furthermore, the recent observation by the
PHENIX Collaboration \cite{PHENIX1} that the elliptic flow $v_2(p_T)
$ of 'direct photons' produced in minimal bias Au+Au collisions at
$\sqrt{s_{NN}}=200$~GeV is comparable to that of the produced pions
was a surprise and in contrast to the theoretical expectations and
predictions. Indeed, the photons produced by partonic interactions
in the quark-gluon plasma phase have not been expected to show a
considerable flow because - in a hydrodynamical picture - they are
dominated by the emission at high temperatures, i.e. in the initial
phase before the elliptic flow fully develops. Since the direct
photon $v_2(\gamma^{dir})$ is a 'weighted average' ($w_i$) of the
elliptic flow of individual contributions $i$
\begin{eqnarray}
\label{v2dir}
 v_2 (\gamma^{dir}) = \sum _i  v_2 (\gamma^{i})
w_i =  \frac{\sum _i  v_2 (\gamma^{i}) N_i }{\sum_i N_i },
\end{eqnarray}
a large QGP contribution gives a smaller $v_2(\gamma^{dir})$. A
sizable photon $v_2$ has been observed also by the ALICE
Collaboration in Pb+Pb collisions at the LHC \cite{ALICE_v2}. None
of the theoretical models could describe simultaneously the photon
spectra and $v_2$ which may be noted as a 'puzzle' for theory.
Moreover, the PHENIX and ALICE Collaborations have reported recently
the observation of non-zero triangular flow $v_3$ (see
\cite{RuanQM14,BockQM14}). Thus, the consistent description of the
photon experimental data remains a challenge for theory.

\subsection{Transport analysis of the photon $v_2$ 'puzzle'}

It is important to stress that  state-of-the art hydro models
reproduce well the hadronic 'bulk' observables (e.g. rapidity
distributions, $p_T$ spectra and $v_2, v_3$ of hadrons). However, in
spite of definite improvements of the general dynamics by including
the fluctuating initial conditions (IP-Glasma or MC-Glauber type)
and viscous effects, the hydro models underestimate the spectra and
$v_2$ of photons at RHIC and LHC energies. For a recent overview we
refer the reader to Ref. \cite{UHeinz}.

As a 'laboratory' for a detailed theoretical analysis we use the
microscopic Parton-Hadron-String Dynamics (PHSD) transport approach
\cite{PHSD}, which is based on the generalized off-shell transport
equations derived in first order gradient expansion of the
Kadanoff-Baym equations, and applicable for strongly interacting
systems. The approach consistently describes the full evolution of a
relativistic heavy-ion collision from the initial hard scatterings
and string formation through the dynamical deconfinement phase
transition to the strongly-interacting quark-gluon plasma  as well
as dynamical hadronization and the subsequent interactions in the
expanding hadronic phase as in the HSD transport approach
\cite{CBRep98}. The partonic dynamics is based on the Dynamical
Quasi-Particle Model (DQPM), that is constructed to reproduce
lattice QCD (lQCD) results for a quark-gluon plasma in thermodynamic
equilibrium. The DQPM provides the mean felds for gluons/quarks and
their effective 2-body interactions that are implemented in the PHSD
(for the details see Ref.~\cite{Cassing:2008nn} and
\cite{Linnyk:Photon,PHSD}). The PHSD model reproduces a large
variety of observables from SPS to LHC energies, e.g. transverse
mass and rapidity spectra of charged hadrons, dilepton spectra,
collective flow coefficients etc. \cite{PHSD,Linnyk:Photon}. Since
the QGP radiation in PHSD occurs from the massive off-shell
quasi-particles with spectral functions, the corresponding QGP rate
has been extended beyond the standard pQCD  rate \cite{AMY01} - see
Ref. \cite{Linnyk11}.

\begin{figure}
\begin{center}
\includegraphics*[width=0.85\textwidth]{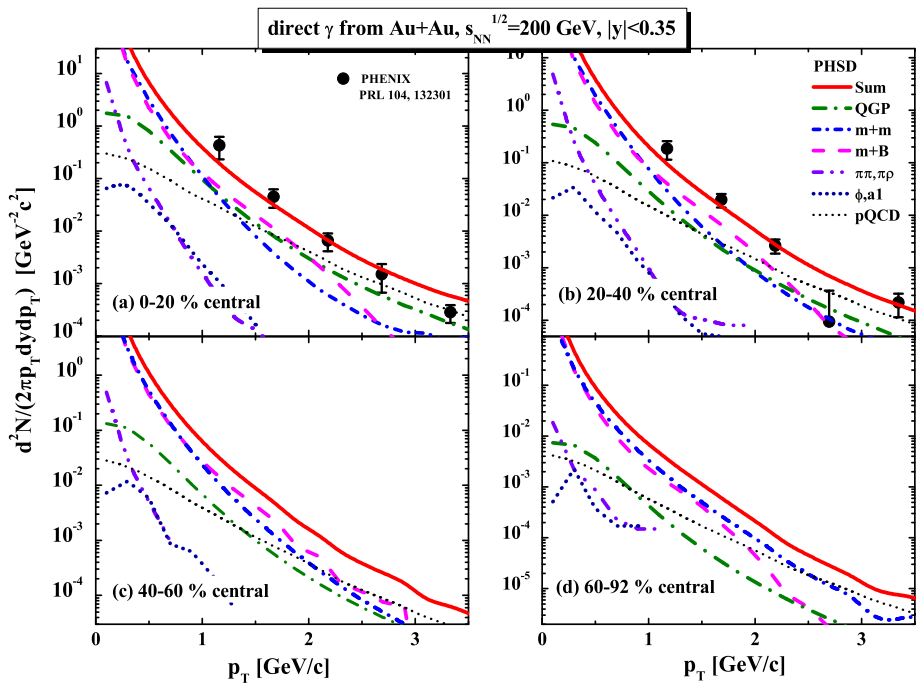}
\caption{Direct photon $p_T$-spectrum from the PHSD approach in
comparison to the PHENIX data \cite{PHENIX2010} at midrapidity for
different centralities in Au+Au collisions at $\sqrt{s_{NN}}=200$
GeV. The channel description is given in the legend. The figure is
taken from Ref. \cite{Linnyk:Photon}. } \label{fig1}
\end{center}
\end{figure}

\begin{figure}
\hspace{0.2cm}
\includegraphics*[width=0.4\textwidth]{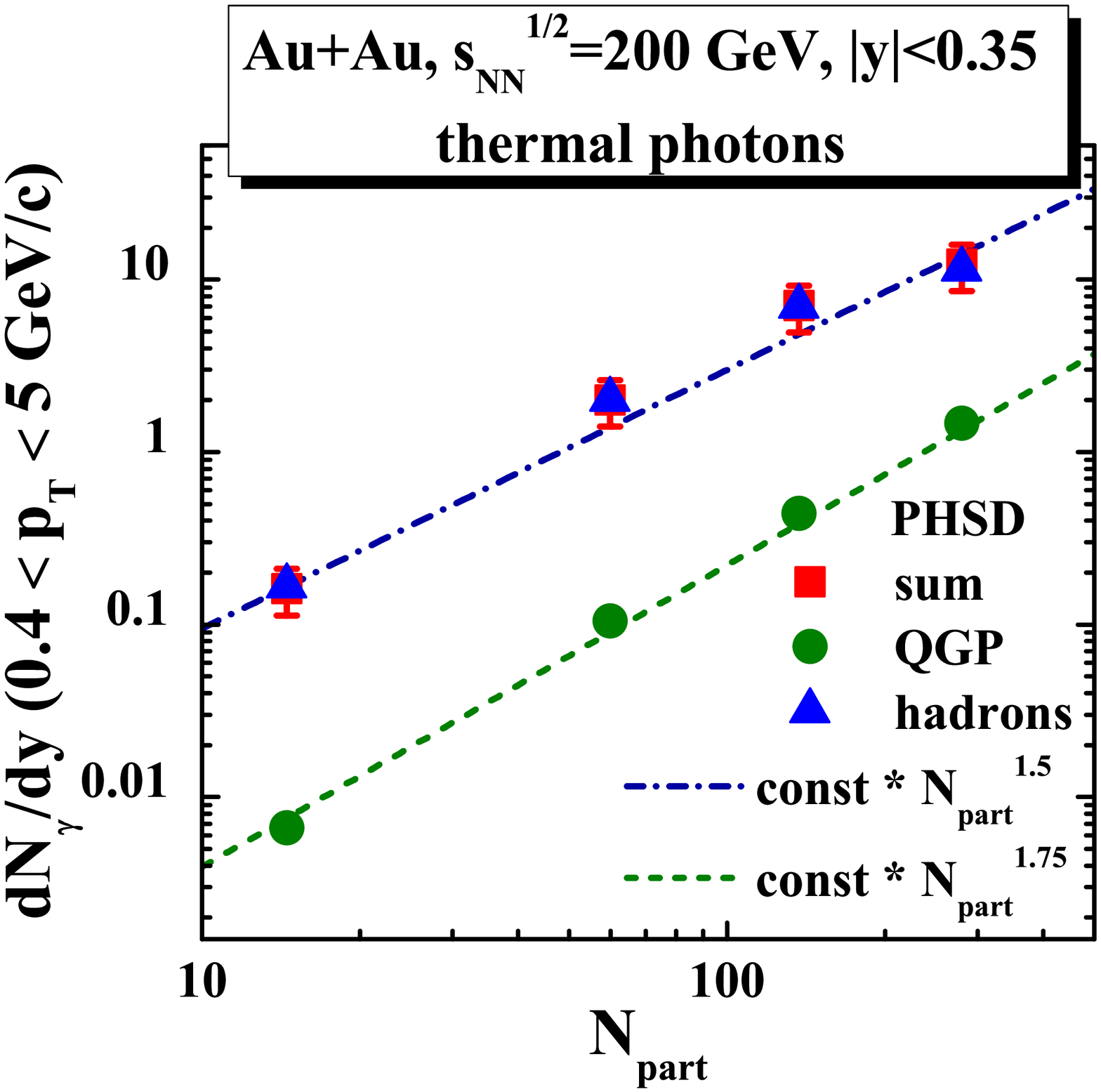}\includegraphics*[width=0.5 \textwidth]{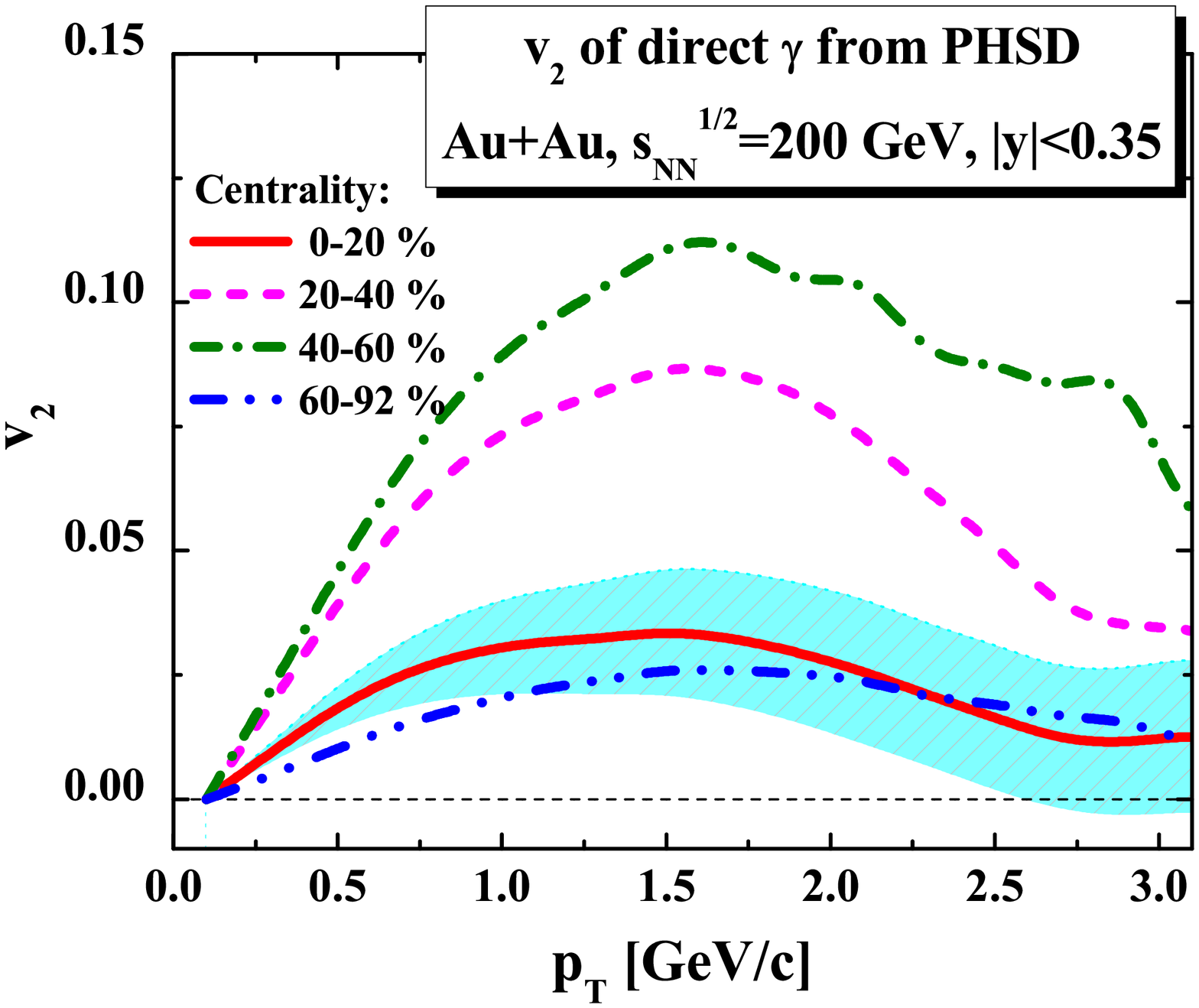}
\caption{(l.h.s.) Integrated spectra of thermal photons produced in
Au + Au collisions at   $\sqrt{s_{NN}}=200$ GeV versus the number of
participants $N_{part}$. The scaling with $N_{part}$ from the QGP
contributions (full dots) and the bresstrahlungs channels (full
triangles) are shown separately. (r.h.s.) The elliptic flow
$v_2(p_T)$ of direct photons produced by binary processes  in Au +
Au collisions at   $\sqrt{s_{NN}}=200$ GeV  for different
centralities versus the photon transverse momentum $p_T$. The
hatched area displays the statistical uncertainty (for the most
central bin). } \label{fig2}
\end{figure}

The result of the PHSD approach \cite{Linnyk:Photon} for the direct
photon $p_T$-spectrum  at midrapidity for Au+Au collisions at
$\sqrt{s}=200$ GeV is shown in Fig. \ref{fig1} for different
centralities in comparison to the PHENIX data \cite{PHENIX2010}. The
upper solid lines give the total direct photon spectra whereas the
various lines show the contributions from indidividual channels (see
legend). While the 'hard' $p_T$ spectra are dominated by the
'prompt' (pQCD) photons, the 'soft' spectra are filled by the
'thermal' sources: the QGP gives up to $~50\%$ of the direct photon
yield between 1 and 2 GeV/$c$ for the most central bin (0-20 \%), a
sizable contribution stems from hadronic sources such as meson-meson
($mm$) and meson-Baryon ($mB$) bremsstrahlung; the contributions
from binary $mm$ reactions are of subleading order. Thus, according
to the present PHSD results the $mm$ and $mB$ bremsstrahlung turn
out to be an important source of direct photons. We note, that the
bremsstrahlung channels are not included in the $mm$ binary 'HG'
rate \cite{TRG04} used in the hydro calculations mentioned above. We
stress, that $mm$ and $mB$ bremsstrahlung cannot be subtracted
experimentally from the photon spectra and has to be included in
theoretical considerations. As has been pointed out earlier its
importance for 'soft' photons follows also from the WA98 data at
$\sqrt{s}=17.3$ GeV \cite{HSD08,LiuRapp07}.

However, some words of caution have to be given here related to the
uncertainties in the  bremsstrahlung channels in the present PHSD
results. The implementation of photon bremsstrahlung from hadronic
reactions in transport approaches \cite{HSD08,Linnyk:Photon} is
based on the 'soft-photon' approximation (SPA) \cite{Rate2} which
implies the factorization of the amplitude for the $a+b\to
a+b+\gamma$ processes to the strong and electromagnetic parts
assuming that the radiation from internal lines is negligible and
the strong interaction vertex is on-shell. In this case the strong
interaction part can be approximated by the on-shell elastic cross
section for the reaction $a+b \to a+b$. Thus, the resulting yield of
the bremsstrahlung photons depends on the validity of the SPA for
large $p_T$ itself and assumptions on the cross sections  for the
meson-meson and meson-baryon elastic scattering which are little (or
not at all) known experimentally. For a more detailed discussion on
uncertainties we refer the reader to Ref. \cite{Linnyk:Photon}. In
this respect we consider the PHSD results on bremsstrahlung as an
'upper estimate'.

The question: "what dominates the photon spectra - {\it QGP
radiation or hadronic contributions}" can be addressed
experimentally by investigating the centrality dependence of the
photon yield: the QGP contribution is expected to decrease when
going from central to peripheral collisions where the hadronic
channels should be dominant. The centrality dependence of the direct
photon yield, integrated over different $p_T$ ranges, has been
measured by the PHENIX Collaboration, too
\cite{PHENIXcd14,MizunoQM14}. It has been found that the midrapidity
'thermal' photon yield scales with the number of participants as
$dN/dy \sim N_{part}^\alpha$ with $\alpha =1.48\pm 0.08$ and only
very slightly depends on the selected $p_T$ range (which is still in
the 'soft' sector, i.e. $< 1.4$ GeV/$c$). Note that the 'prompt'
photon contribution (which scales as the $pp$ 'prompt' yield times
the number of binary collisions in $A+A$) has been subtracted from
the data. The PHSD predictions \cite{Linnyk:Photon} for  Au+Au
collisions at different centralities give $\alpha (total) \approx
1.5$, which is dominated by hadronic contributions, while the QGP
channels scale with $\alpha (QGP) \sim 1.7$ (see Fig. 2 (l.h.s.)). A
similar finding has been obtained by the viscous (2+1)D VISH2+1 and
(3+1)D MUSIC hydro models \cite{VISH_McG}: $\alpha(HG) \sim 1.46, \
\ \alpha(QGP) \sim 2, \ \ \alpha(total) \sim 1.7$. Thus, the QGP
photons show a centrality dependence significantly stronger than
that of hadron-gas (HG) photons.

In Fig.~\ref{fig2} (r.h.s.) we provide predictions for the
centrality dependence of the direct photon elliptic flow $v_2(p_T)$
within the PHSD approach. The direct photon $v_2$ is seen to be
larger in the peripheral collisions compared to the most central
ones. The predicted centrality dependence of the direct photon flow
results from the interplay of two independent factors: First, the
channel decomposition of the direct photon yield  changes (cf. Fig.
1): the admixture of photons from the hadronic phase increases for
more peripheral collisions. Since the PHSD approach predicts a very
small $v_2$ of photons produced in the initial hot deconfined phase
by partonic channels  of the order of 2\% the photon flow $v_2$
shows about the same signal for the most central bin. On the other
hand, the photons from the hadronic sources show a strong elliptic
flow (up to 10\%), on the level of the $v_2$ of final hadrons.
Accordingly, since the channel decomposition of the direct photons
changes with centrality, the elliptic flow of the direct photons
increases with decreasing centrality and becomes roughly comparable
with the elliptic flow of pions in peripheral collisions. The
elliptic flow in the most peripheral bin is also low in
Fig.~~\ref{fig2} (r.h.s.) because all the colliding particles have
little flow at this high impact parameter $b$.

%----------------------------------------------------------------------------------
\section{Dileptons}

\subsection{Production sources}
Dileptons ($e^+e^-$ or $\mu^+\mu^-$ pairs) can be emitted from all
stages of the reactions as well as a photons. One of the advantages
of dileptons compared to photons is an additional 'degree of
freedom' - the invariant mass $M$ which allows to disentangle
various sources. The following production sources of dileptons in
$p+p, p+A$ and $A+A$ collisions are leading:\\
1) Hadronic sources:\\
(i) at low invariant masses ($M < 1$ GeV$c$) -- the  Dalitz decays of mesons and
 baryons $(\pi^0,\eta,\Delta, ...)$ and the direct decay of
vector mesons  $(\rho, \omega, \phi)$ as well as hadronic bremsstrahlung; \\
(ii) at intermediate masses ($1< M < 3$ GeV$c$) --
leptons from correlated $D+\bar D$ pairs, radiation
from multi-meson reactions
($\pi+\pi, \ \pi+\rho, \ \pi+\omega, \ \rho+\rho, \ \pi+a_1, ... $)  -
so called $'4\pi'$ contributions; \\
(iii) at high invariant masses ($M > 3$ GeV$c$) -- the direct decay of
vector mesons  $(J/\Psi, \Psi^\prime)$ and
initial 'hard' Drell-Yan annihilation to dileptons
($q+\bar q \to l^+ +l^-$, where $l=e,\mu$).\\
2) 'thermal' QGP dileptons radiated from the partonic interactions
in heavy-ion ($A+A$) collisions that contribute dominantly to
the intermediate masses. The leading processes
are the 'thermal' $q\bar q$ annihilation ($q+\bar q \to l^+ +l^-$, \ \
$q+\bar q \to g+ l^+ +l^-$) and Compton scattering
($q(\bar q) + g \to q(\bar q) + l^+ +l^-$).

\subsection{Transport results from SIS to LHC energies}
% SIS -----------------------
 At energies around 1 $A$GeV dileptons have been measured in
heavy-ion collisions at the BEVALAC  in Berkeley by the DLS
Collaboration by more than two decades ago. These data led to the so
called 'DLS puzzle' because the DLS dilepton yield   in C+C and
Ca+Ca collisions at 1 $A$GeV in the invariant mass range from 0.2 to
0.5 GeV  was about five times higher than the results from different
transport models at that time using the 'conventional' dilepton
sources such as bremsstrahlung, $\pi^0, \eta, \omega$ and $\Delta$
Dalitz decays and direct decay of vector mesons ($\rho, \omega,
\phi$) \cite{Bratkovskaya:1996bv}.   To solve this puzzle was one of
the main motivations to build  the HADES (High Acceptance Dilepton
Spectrometer) detector at  GSI \cite{Agakishiev:2011vf}.
Indeed the HADES Collaboration could confirm the DLS measurements at
1  $A$GeV when passing their events for C+C through the DLS filter
\cite{xxhades}. From the theory side it was argued that the pn
bremsstrahlung channel should be sizeably enhanced as compared to
the early soft photon calculations  \cite{Bratkovskaya:1996bv}.
Indeed, a good reproduction of various spectra a different energies
could be achieved within the HSD calculations in Ref.
\cite{xxbraca}. Note, however, that even the bremsstrahlung from pn
reactions at these low energies is discussed controversally in the
community and not available experimentally.  We here report on the
actual status of the transport calculations in comparison to the
HADES data \cite{BratAich}.
\begin{figure}[th!]
%\phantom{a}\vspace*{5mm}
\includegraphics*[width=6.8cm]{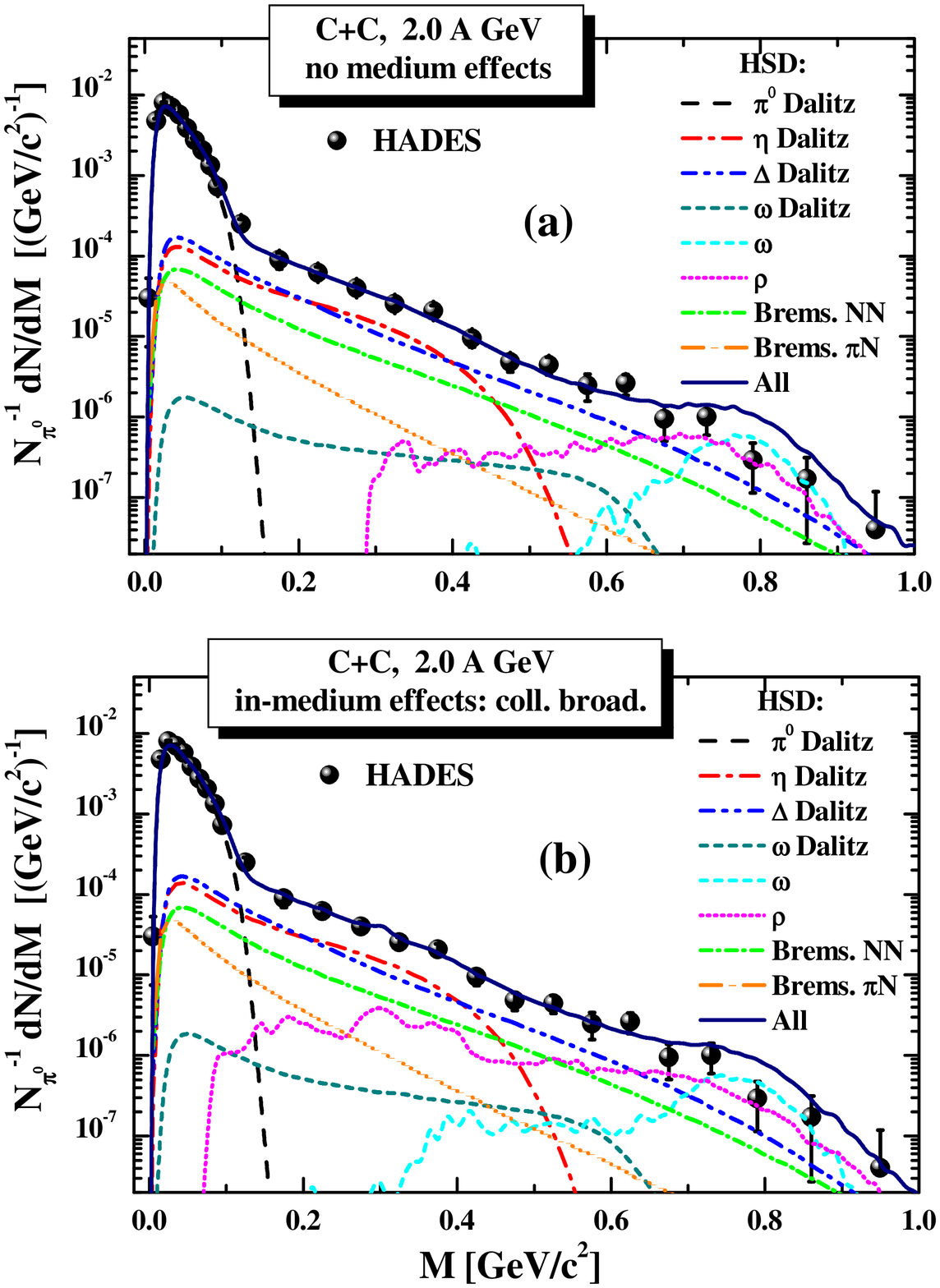}\includegraphics*[width=6.8cm]{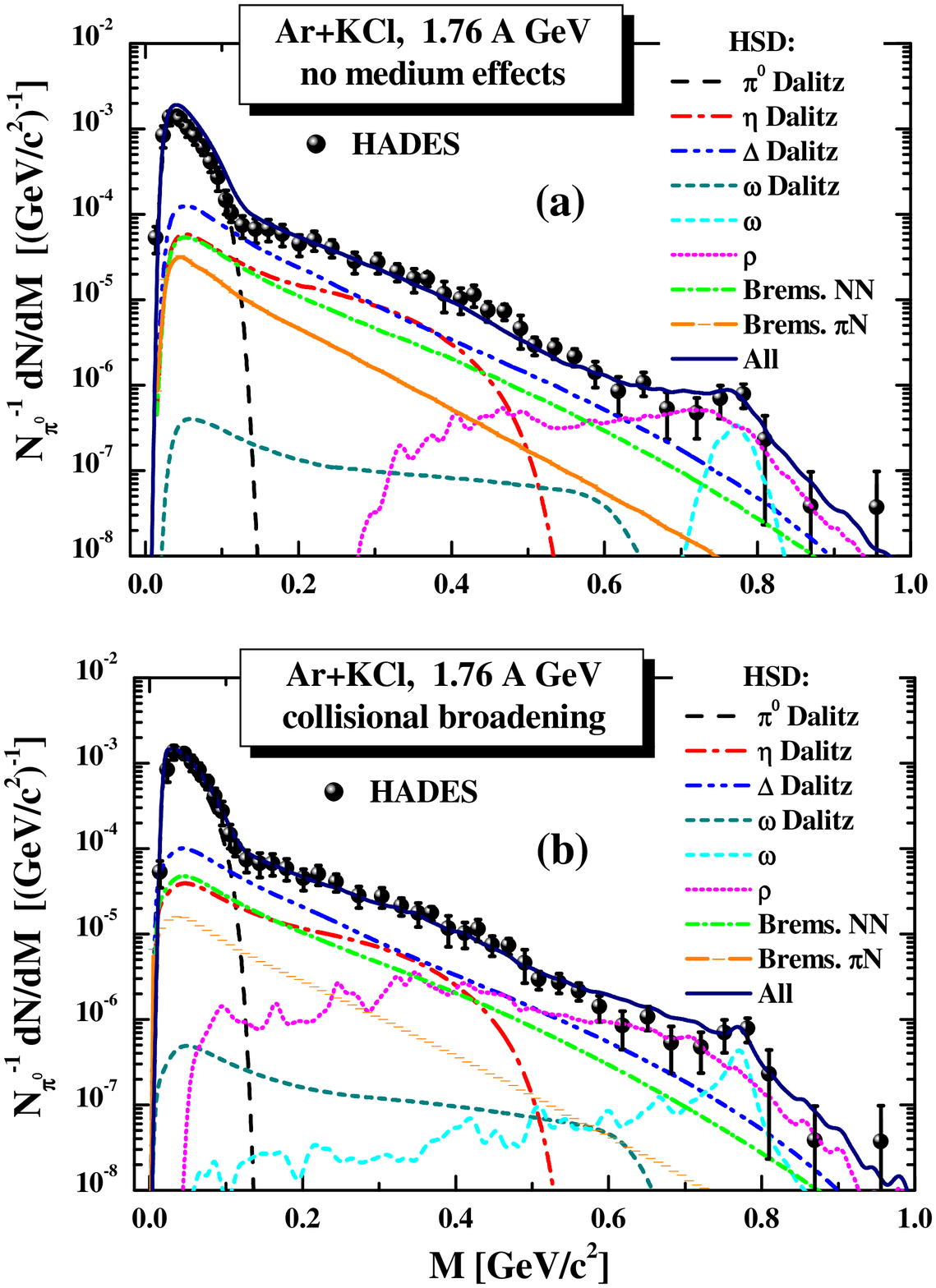}
\vspace*{5mm}
\caption{The mass differential dilepton spectra - normalized to the
$\pi^0$ multiplicity
 - from HSD calculations for C+C  at 2 $A$GeV (l.h.s.) and Ar+KCl
 at 1.76 $A$GeV (r.h.s.) in
comparison to the HADES data
\cite{Agakishiev:2009yf,Agakishiev:2011vf}.  The upper parts (a)
shows the case of 'free' vector-meson spectral functions while the
lower parts (b) give the result for the 'collisional broadening'
scenario. The different colour lines display individual channels in
the transport calculation (see legend).  The theoretical
calculations passed through the corresponding HADES acceptance
filter and mass/momentum resolutions. } \label{Fig_CC20}
\end{figure}

Fig. \ref{Fig_CC20} (l.h.s.) shows the mass differential dilepton
spectra - normalized to the $\pi^0$ multiplicity - from HSD
calculations for C+C at 2 $A$GeV in comparison to the HADES data
\cite{Agakishiev:2009yf}. The theoretical calculations passed
through the corresponding HADES acceptance filters and mass/momentum
resolutions  which leads to a smearing of the spectra at high
invariant mass and particularly in the $\omega$ peak region. The
upper part shows  the case of 'free' vector-meson spectral functions
while the lower part presents the result for the 'collisional $\rho$
broadening' scenario. Here the difference between  in-medium
scenarios is of minor importance and partly due to the limited mass
resolution which smears out the spectra. Fig. \ref{Fig_CC20}
(r.h.s.) displays the mass differential dilepton spectra -
normalized to the $\pi^0$multiplicity - from HSD calculations for
the heavier system Ar+KCl at 1.76 $A$GeV  in comparison to the HADES
data \cite{Agakishiev:2011vf}.  The upper part shows again the case
of 'free' vector-meson spectral functions while the lower part gives
the result for the 'collisional broadening' scenario. Also in this
data set the enhancement around the $\rho$ mass is clearly visible.
For the heavier system the 'collisional broadening' scenario shows a
slightly better agreement with experiment than the 'free' result and
we expect that for larger systems the difference between the two
approaches increases. We note that with increasing mass A+A of the
system the low mass dilepton regime from roughly 0.15 to 0.5 GeV
increases due to multiple $\Delta$-resonance production and Dalitz
decay. The dileptons from intermediate $\Delta$'s, which are part of
the reaction cycles $\Delta \to \pi N ; \pi N \to \Delta$ and $NN\to
N\Delta; N\Delta \to NN$, escape from the system while the decay
pions do not \cite{BratAich}.   With increasing system size more generations of
intermediate $\Delta$'s are created and the dilepton yield enhanced
accordingly. In inclusive C+C collisions there is only a moderate
enhancement relative to scaled p+p and p+n collisions due to the
small size of the system while in Ar+KCl reactions already several
(3-4) reactions cycles become visible. On the other hand the effects
from a broadened vector-meson spectral function is barely visible
for both systems and calls for Au+Au collisions. Indeed, the
respective data have been taken and are currently analyzed. For
detailed predictions we refer the reader to Ref. \cite{BratAich}.

%---- SPS ---
Dileptons from heavy-ion collisions  at SPS energies have been
measured in the last decades by the CERES \cite{CERES} and NA60
\cite{NA60} Collaborations. The high accuracy dimuon NA60 data
provide a unique possibility to subtract the hadronic cocktail from
the spectra and to distinguish  different in-medium scenarios for
the $\rho$-meson spectral function such as a collisional broadening
and dropping mass \cite{ChSym,Li:1995qm}. The main messages obtained
by a comparison of the variety of model calculations (see e.g.
\cite{ChSym,dilHSD,dilSPStheor}) with experimental data can be
summarized as \\ (i) the low mass spectra \cite{CERES,NA60} provide
a clear evidence for the collisional broadening of the $\rho$-meson
spectral function in the hot and dense medium; \\ (ii) the
intermediate mass spectra above $M>1$ GeV/$c^2$ \cite{NA60} are
dominated by partonic radiation; \\ (iii) the rise and fall of the
inverse slope parameter of the dilepton $p_T$-spectra  (effective
temperature) $T_{eff}$ \cite{NA60} provide evidence for the thermal
QGP radiation; \\ (iv) isotropic angular distributions \cite{NA60}
are an indication for a thermal origin of dimuons. \\

%--- PHENIX ----
An increase in energy from SPS to RHIC has opened  new possibilities
to probe by dileptons a possibly different matter at very high
temperature, i.e. dominantly in the QGP stage, created in central
heavy-ion collisions. The dileptons ($e^+e^-$ pairs) have been
measured first by the PHENIX Collaboration for $pp$ and $Au+Au$
collisions at $\sqrt{s}=200$ GeV \cite{PHENIXdil}. A large
enhancement of the dilepton yield relative to the scaled $pp$
collisions in the invariant mass regime from 0.15 to 0.6 GeV/$c^2$
has been reported for central Au+Au reactions. This observation has
stimulated a lot of theoretical activity (see the model comparison
with the data in Ref. \cite{PHENIXdil}). The main messages - which hold
up-to-now - can be condensed such that the theoretical models, which
provide a good description of $pp$ dilepton data and peripheral
$Au+Au$ data, fail in describing the excess in central collisions
even with in-medium scenarios for the vector-meson spectral
function \cite{dilHSD}. The missing strengths might be attributed to low $p_T$
sources \cite{dilPHSDRHIC}. On the other hand the intermediate mass
spectra are dominated by the QGP radiation as well as leptons from
correlated charm pairs ($D+\bar D$) \cite{dilHSD,dilPHSDRHIC,Rapp13}.

% --- STAR ------

\begin{figure}[t]
\hspace{1cm}
\includegraphics*[width=0.9\textwidth]{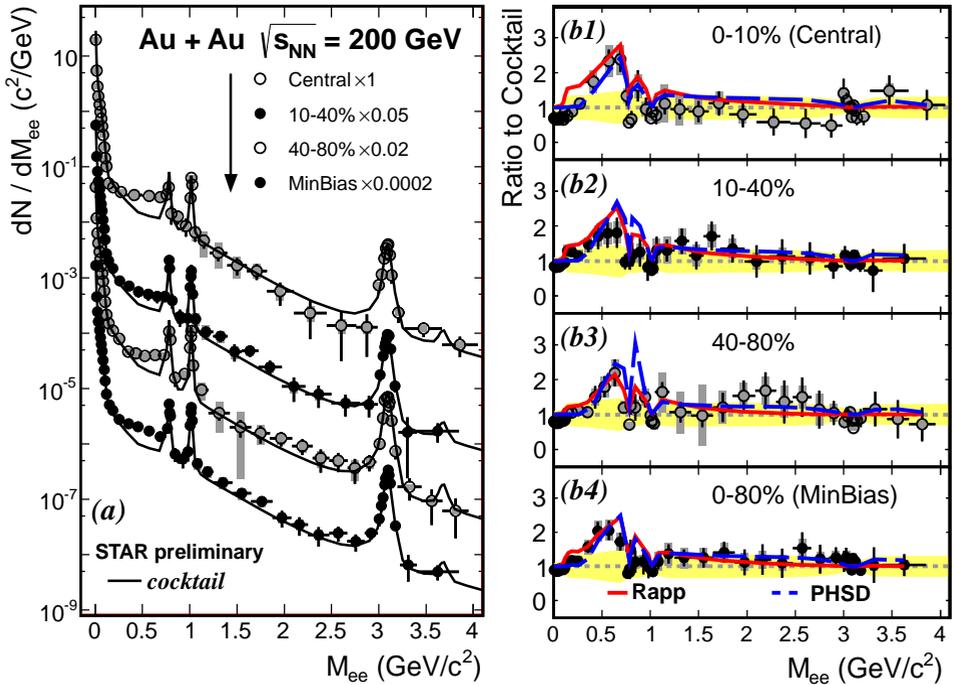}
\caption{Centrality dependence of the midrapidity dilepton yields
(left) and its ratios (right) to the 'cocktail' for 0-10\%, 10-40\%,
40-80\%, 0-80\% central Au+Au collisions at $\sqrt{s}=200$ GeV: a
comparison of STAR data with theoretical predictions from the PHSD
('PHSD' - dashed lines) and the expanding fireball model ('Rapp' -
solid lines). The figure is taken from Ref. \cite{HuckQM14}.}
\label{fig:dilSTAR200cd}
\end{figure}

In this respect it is very important to have independent
measurements which have been carried out by the STAR Collaboration
\cite{dilSTAR}. Fig. \ref{fig:dilSTAR200cd} shows the comparison of
STAR data of midrapidity dilepton yields (l.h.s.) and its ratios
(r.h.s.) to the 'cocktail' for  0-10\%, 10-40\%, 40-80\%, 0-80\%
central Au+Au collisions at $\sqrt{s_{NN}}=200$ GeV  in comparison
to the theoretical model predictions from the PHSD approach and the
expanding fireball model of Rapp and collaborators. As seen from
Fig. \ref{fig:dilSTAR200cd} the excess of the dilepton yield over
the expected cocktail is larger for very central collisions and
consistent with the model predictions including the collisional
broadening of the $\rho$-meson spectral function at low invariant
mass and QGP dominated radiations at intermediate masses. Moreover,
the recent STAR dilepton data for Au+Au collisions from the Beam
Energy Scan (BES) program  for  $\sqrt{s_{NN}}=19.6, 27, 39$ and 62.4
GeV \cite{RuanQM14,dilSTAR2,HuckQM14} are also in line with the
expanding fireball model (as well as PHSD) predictions with a $\rho$
collisional broadening \cite{HuckQM14}. According to the PHSD
calculations the excess is increasing with decreasing energy due to
a longer $\rho$-propagation in the high baryon density phase (see
Fig. 3 in \cite{RuanQM14}).

\begin{figure}[t]
\hspace{0.1cm}
\includegraphics[width=0.45\textwidth]{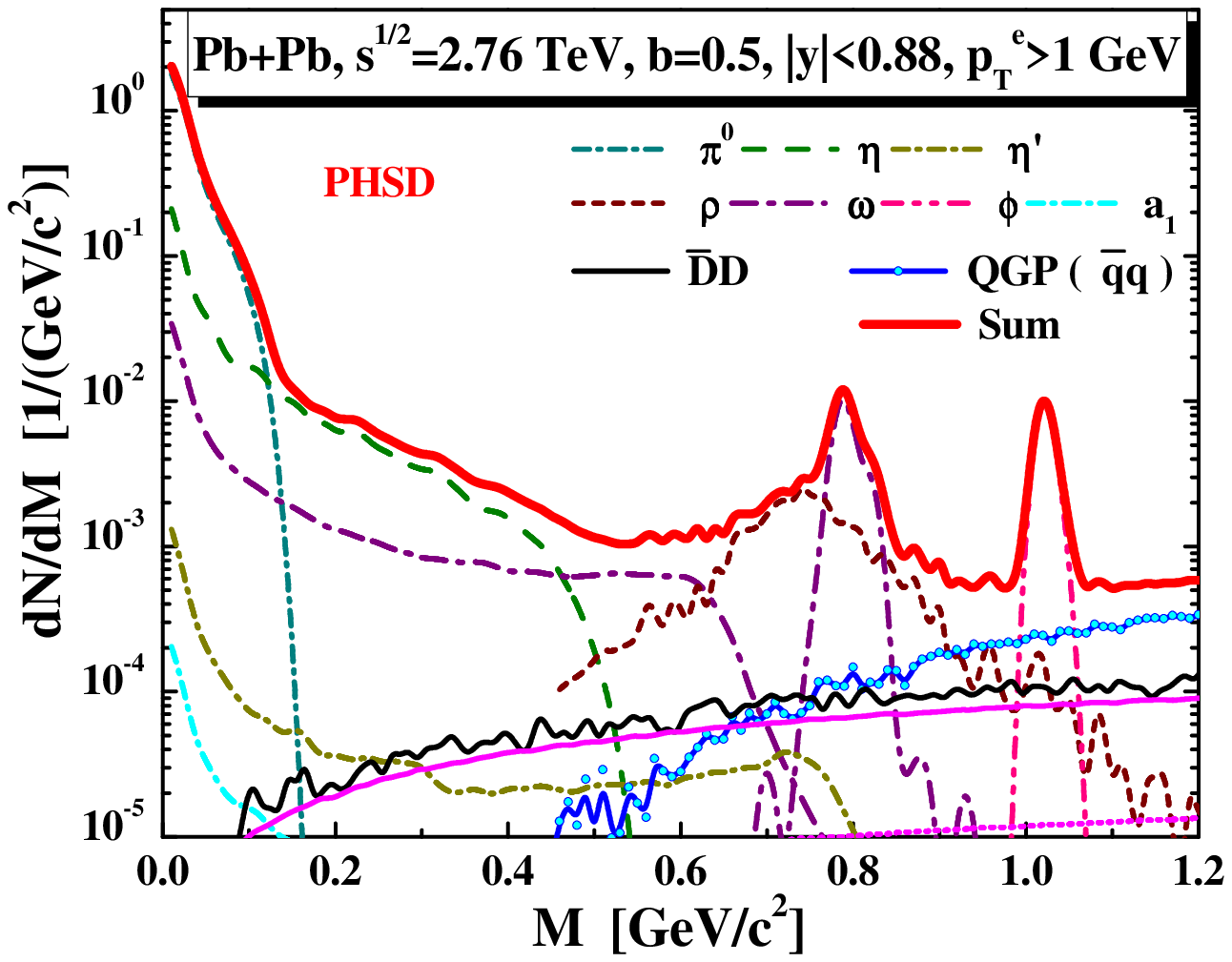}\includegraphics[width=0.48\textwidth]{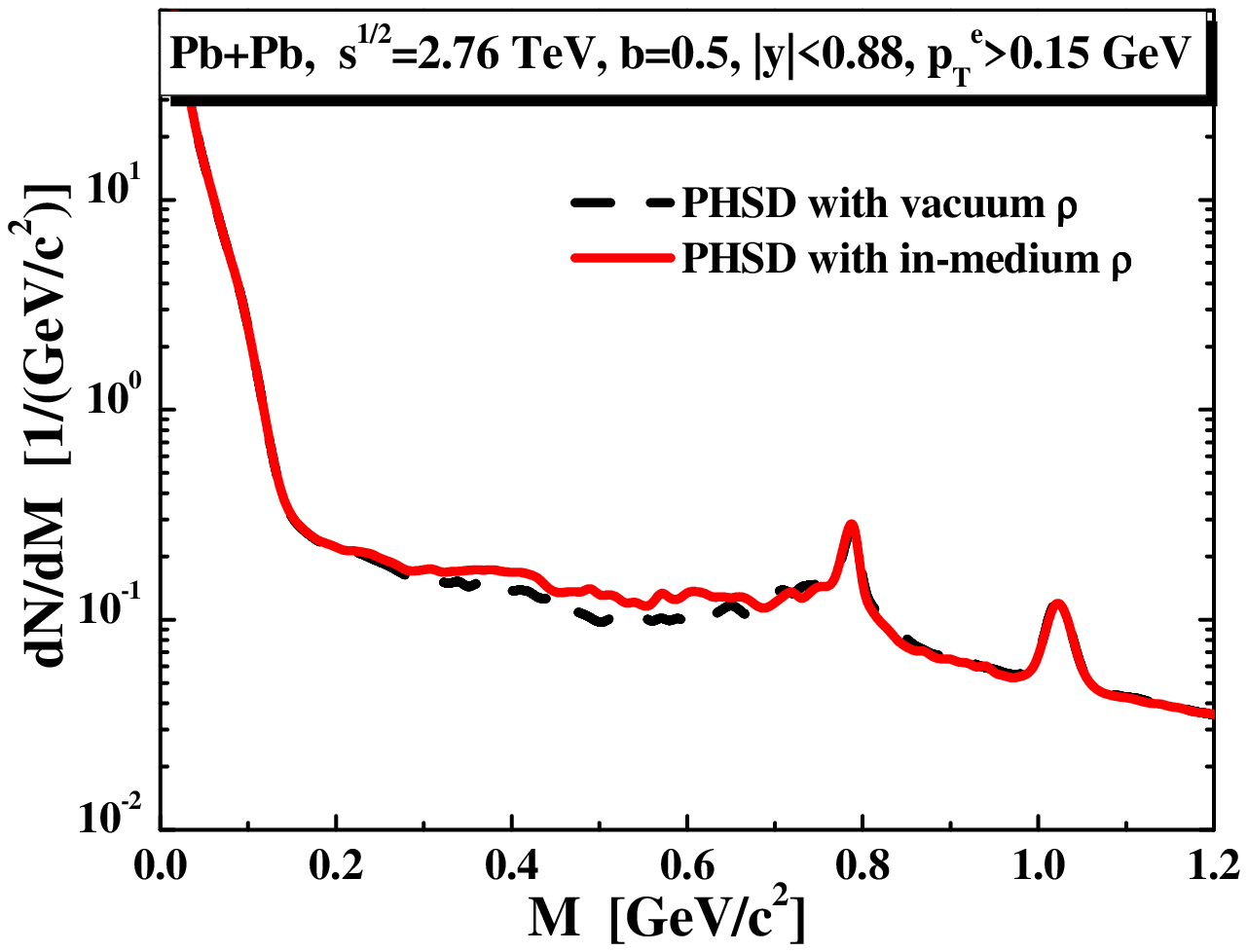}
\vspace{4mm}
\caption{Midrapidity dilepton yields for Pb+Pb at $\sqrt{s_{NN}}$ =
2.76 TeV (l.h.s.) for a lepton $p_T$ cut of 1 GeV/c. The channel
decomposition is explained in the legend. (r.h.s.) Same as for the
l.h.s. but for a lepton $p_T$ cut of 0.15 GeV/c for a 'free' $\rho$
spectral function (dashed line) and the collisional broadening
scenario (solid line). The figures are taken from Ref.
\cite{dilPHSDLHC}.} \label{fig5}
\end{figure}

The upcoming PHENIX data for central
Au+Au collisions - obtained after an upgrade of the detector -
together with the BES-II RHIC data should provide finally a
consistent picture on the low mass dilepton excess in relativistic
heavy-ion collisions.
On the other hand, the upcoming ALICE data \cite{dilALICE} for
heavy-ion dileptons for Pb+Pb at $\sqrt{s}$ = 2.76 TeV will give a
clean access to the dileptons emitted from the QGP
\cite{Rapp13,dilPHSDLHC}. In Fig. \ref{fig5} (l.h.s.)  we present
the PHSD predictions for central Pb+Pb collisions \cite{dilPHSDLHC}
in the low mass sector for a realistic lepton $p_T$ cut of 1 GeV/c. It is
clearly seen that the QGP sources and contribution from correlated
$D{\bar D}$ pairs are subleading in the low mass regime where we find
the conventional hadronic sources. For a lepton $p_T$ cut of 1 GeV/c (l.h.s.)
one practically cannot identify an effect of the $\rho$ collisional
broadening in the dilepton spectra in the PHSD calculations. Only
when applying a low $p_T$ cut of 0.15 GeV/c a small enhancement of
the dilepton yield from 0.3 to 0.7 GeV becomes visible (r.h.s. of
Fig. \ref{fig5}). This low sensitivity to hadronic in-medium effects
at LHC energies is due to the fact that the hadrons come out late in central Pb+Pb
collisions and are boosted to high velocities due to the high pressure in
the early partonic phase.

% --- v2, v3 -----
In the end, we mention that promising perspectives with dileptons
have been suggested in Ref. \cite{v3dil} to measure the anisotopy
coefficients  $v_n, \ n=2,3$ similar to photons. The calculations
with the viscous (3+1)d MUSIC hydro for central Au+Au collisions at
RHIC energies show that $v_2, v_3$ are sensitive to the dilepton
sources and to the EoS and $\eta/s$ ratio. The main advantage of
measuring flow coefficients $v_n$ with dileptons compared to photons
is the fact that an extra degree of freedom $M$ might allow to
disentangle the sources.

%----------------------------------------------------------------------------------
\section{Conclusions}
In conclusion, our calculations show that the photon production in
the QGP is dominated by the early phase (similar to hydrodynamic
models) and is localized in the center of the fireball, where the
collective flow is still rather low, i.e. on the 2-3 \% level, only.
Thus, the strong $v_2$ of direct photons - which is comparable to
the hadronic $v_2$ - in PHSD is attributed to hadronic channels,
i.e. to meson binary reactions, meson-meson and meson-baryon bremsstrahlung
which are not subtracted in the data. On the other hand, the strong $v_2$ of the
'parent' hadrons, in turn, stems from the interactions in the QGP.
We have argued that a precise measurement of the centrality dependence of the
elliptic flow of direct photons together with their differential spectra should
help in clarifying the the photon $v_2(p_T)$ "puzzle". Note however, that the
hadronic bremsstrahlungs channels are not well under control and our present
results should be taken as 'upper limits'. Some more work will have to be done
in this direction.

The main messages from our dilepton campaign may be formulated as follows: i) the
low mass ($M=0.2-0.6$ GeV/$c^2$) dilepton spectra show sizable changes due to
hadronic in-medium effects, i.e. multiple hadronic resonance formation (at SIS energies)
or a modification of the properties of
vector mesons  (such as collisional broadening) in the hot and dense
hadronic medium (partially related to chiral symmetry
restoration); these effects can be observed at all energies up to LHC
(preferentially in heavy systems) but are most pronounced in the FAIR/NICA energy regime; (ii) at
intermediate masses the QGP ($q\bar q$ thermal radiation) dominates
for $M>1.2$ GeV/$c^2$, it grows with
increasing energy and becomes dominant at the LHC energies.
The dilepton measurements within the future experimental energy and
system size scan ($pp, pA, AA$) from low to top RHIC energies as well as
new ALICE data at LHC energies will extend our knowledge on the
properties of hadronic and partonic matter via its electromagnetic
radiation.

\vspace*{2mm} %\noindent
The authors acknowledge financial support through the 'HIC for FAIR'
framework of the 'LOEWE' program and like to thank all their
coauthors for their help and valuable contributions.

\end{document}